\documentclass[a4paper,fleqn]{cas-dc}

\usepackage[numbers]{natbib}
\usepackage{amsmath,amssymb}
\usepackage{latexsym}
\usepackage{subfig}

\begin{document}

\shortauthors{Peyman Tahghighi et~al.}

\title [mode = title]{Analysis of Macula on Color Fundus Images Using Heightmap Reconstruction Through Deep Learning}

\author[1]{Peyman Tahghighi}
\ead{ peyman.tahghighi@ut.ac.ir}
\cortext[cor1]{Corresponding author}
\author[1]{Reza A.Zoroofi}
\author[2]{Sare Safi}
\author[3,4]{Alireza Ramezani}
\address[1]{School of Electrical and Computer Engineering,University of Tehran,Tehran, Iran}
\address[2]{Ophthalmic Research Center, Research Institute for Ophthalmology and Vision Science\\ Shahid Beheshti University of Medical Sciences, Tehran, Iran}
\address[3]{Ophthalmic Epidemiology Research Center, Research Institute for Ophthalmology and Vision Science\\ Shahid Beheshti University of Medical Sciences, Tehran, Iran}
\address[4]{Negah Specialty Ophthalmic Research Center\\ Shahid Beheshti University of Medical Sciences, Tehran, Iran}

\begin{abstract}
For medical diagnosis based on retinal images, a clear understanding of 3D structure is often required but due to the 2D nature of images captured, we cannot infer that information. However, by utilizing 3D reconstruction methods, we can recover the height information of the macula area on a fundus image which can be helpful for diagnosis and screening of macular disorders. Recent approaches have used shading information for heightmap prediction but their output was not accurate since they ignored the dependency between nearby pixels and only utilized shading information. Additionally, other methods were dependent on the availability of more than one image of the retina which is not available in practice. In this paper, motivated by the success of Conditional Generative Adversarial Networks(cGANs) and deeply supervised networks, we propose a novel architecture for the generator which enhances the details and the quality of output by progressive refinement and the use of deep supervision to reconstruct the height information of macula on a color fundus image. Comparisons on our own dataset illustrate that the proposed method outperforms all of the state-of-the-art methods in image translation and medical image translation on this particular task. Additionally, perceptual studies also indicate that the proposed method can provide additional information for ophthalmologists for diagnosis.
\end{abstract}

\begin{keywords}
Conditional generative adversarial networks \sep Convolutional neural networks \sep Fundus image \sep Deep learning
\end{keywords}
\maketitle

\section{Introduction}
Vision is crucial for our everyday activities such as driving, watching television, reading and interacting socially and visual impairment can be a real handicap. Even the slightest visual loss can affect our quality of life considerably and cause depression and in old adults cause accidents, injuries and falls \cite{1,2}. Blindness is the final stage of many eye diseases and according to previous studies, the most common cause of visual impairments are cataract, macular degeneration, glaucoma and diabetic retinopathy \cite{3,4}.

\par

Color fundus photography is a 2D imaging modality for the diagnosis of retinal diseases. 3D structure of the eye provides a considerable amount of crucial information (such as information about elevation in different parts) for ophthalmologists to diagnose which is unavailable in 2D fundus images. Therefore, being able to infer this information from just a 2D image can be helpful. Furthermore, the reconstructed heightmap offers clinicians another means to view eye structure which may help them in better and more accurate diagnosis \cite{59,60}. Optical Coherence Tomography (OCT) \cite{61} is an expensive but vital tool for evaluating the retinal structure which provides ophthalmologists with valuable information, enabling them to diagnose most of the macula diseases. Nevertheless, owing to the cost of using this system, OCT devices are not ubiquitous and using fundus images is mostly common for screening.

\par

\begin{figure}[h]
\centering
\includegraphics[clip,trim=5.5cm 19.0cm 3cm 1cm width=1.0\linewidth,scale=0.8]{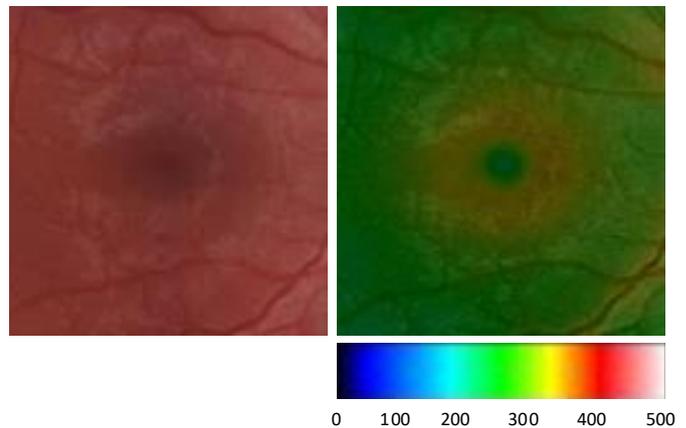}
\caption{The left and right image represent the correspondence between a fundus image and its heightmap image. As can be seen, each pixel's color value of the image on the right indicates a height according to the colorbar which ranges from 0 $\mu m$ to 500 $\mu m$. Note that all numbers are in micrometer.}
\label{fig1}
\end{figure}

Shape from shading (SFS) \cite{70} is the only method applied to this problem for the reconstruction of the height of a single fundus image \cite{60}. However, this method suffers from drawbacks that limits its usage in this particular task. In fact, the success of the SFS method is totally dependent on predicting the position of the light source and the assumption about the surface which cannot be applied to the eye retina. Furthermore, disparity map estimation, which is one of the common methods for 3D reconstruction was also applied to this problem \cite{62}. Nonetheless, it totally depends on the availability of the stereo images from both eyes which is not practical. Hence, devising a method to automatically generate an accurate heightmap image from a given fundus image is crucial.

\par

In recent years, with the advent of Conditional Generative Adversarial Networks (cGANs) \cite{22,23}, many researchers used this methodology for image generation and transformation tasks \cite{24,39,40,76,77,80}. Medical image analysis also benefited a lot from these powerful models and many researchers used these methods for translating between different medical images such as translating between CT and PET images \cite{74,36}, denoise and correct the motion in medical images such as denoising CT images \cite{35} or correcting the motion in MR images \cite{36} and finally segmenting medical images \cite{37,78,79}. Most of these methods benefited from the U-Net architecture \cite{25} which first were proposed for image segmentation and its extension U-Net++ \cite{21}. Perceptual loss \cite{39,40} is another major part of successful methods which considers the difference between two images using high-level features extracted from different layers of a Deep Neural Network (DNN).
\par
Motivated by the promising results of cGANs on tasks relating to the analysis of medical images, in this work, we applied this method to generate a heightmap image from a given color fundus image targeted on the macula area. In fact, height information is one of the crucial information that OCT devices provide to ophthalmologists and is missing in color fundus images due to their 2D nature. Hence, by extracting such information from only a fundus image, we can ease the diagnosis and management of retinal diseases with macular thickness changes.

\par

Considering Figure \ref{fig1}, since our problem can be seen as an image translation task in which we want to predict a color image containing heights data from a fundus image targeted on the macula area, cGANs can be utilized in this problem. That is to say, each pixel in the right image of Figure \ref{fig1} has a color value which represents a height from 0 $\mu m$ to 500 $\mu m$ and by predicting red, green and blue color values for each pixel of the left image (fundus image), we can predict its heightmap. The color bar below Figure \ref{fig1} demonstrates the assignment of different color values to different heights.

\par

In this paper, we used a stack of three U-Nets for our generator network which we averaged on the output of them for deep supervision. Furthermore, in order to avoid problems of traditional GANs, we used Least Squares Adversarial Loss \cite{65} instead along with perceptual loss \cite{39} and L2-loss as pixel reconstruction loss. For the discriminator network, we used an image-level discriminator that classifies the whole image as real or fake. To the best of our knowledge, this is the first research paper on predicting the heightmap of the macula area on fundus images using DNNs. We evaluated our approach qualitatively and quantitatively on our dataset and compared the results with state-of-the-art methods in image translation and medical image translation. Furthermore, we studied the application of our method on real diagnosis cases which showed that reconstructed heightmaps can provide additional information to ophthalmologists and can be used for the analysis of the macula region.

\par

Our main contributions can be listed as follows:
\begin{itemize}
\item Motivated by deeply supervised networks \cite{69}, we propose a novel deep architecture for the generator network based on U-Net \cite{25} and CasNet \cite{36} which consists of a number of connected U-Nets that we utilized each of their output for deep supervision.
\item We propose the first method for the reconstruction of the heightmap of the macula image based on DNNs.
\item Finally, the subjective performance of our reconstructed heightmap was investigated from a medical perspective by two experienced ophthalmologists.
\end{itemize}

\section{Methods}
\subsection{Preprocessing}
A DNN has the capability to learn from unpreprocessed image data, but it can learn more easily and efficiently if we apply appropriate preprocessing on the input data. Hence, in this paper, we first apply Contrast Limited Adaptive Histogram Equalization (CLAHE) \cite{64} which enhances the foreground and background contrast. Afterward, we apply normalization (division by 255) to the input images to prepare them for feeding into the network. The impact of preprocessing on the input fundus images is depicted in Figure \ref{fig2}.

\begin{figure}[h]
\captionsetup[subfigure]{labelformat=empty}
\subfloat[CLAHE]{
\begin{minipage}{
0.23\textwidth}
\includegraphics[width=1\textwidth]{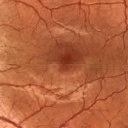}
\end{minipage}}\
\subfloat[Normal]{\begin{minipage}{
0.23\textwidth}
\includegraphics[width=1\textwidth]{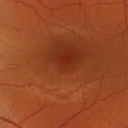}
\end{minipage}}\hspace{0.095cm}
\setcounter{subfigure}{0}
\subfloat[CLAHE]{
\begin{minipage}{
0.23\textwidth}
\includegraphics[width=1\textwidth]{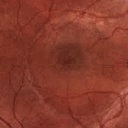}
\end{minipage}}\
\subfloat[Normal]{\begin{minipage}{
0.23\textwidth}
\includegraphics[width=1\textwidth]{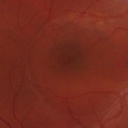}
\end{minipage}}
\caption{The effect of CLAHE preprocessing on the quality of fundus images. As can be seen, in preprocessed images, the details are more clear and this can positively affect learning procedure.}
\label{fig2}

\end{figure}

\subsection{Network structure}
In our proposed cGAN setting, the input to our generator is a 128$\times$128$\times$3 image of the macula area on a fundus image and the generator will generate an image of the same size and depth such that each pixel's color indicates a height as depicted in Figure \ref{fig1}. The discriminator takes this image and gives a probability between 0 to 1 which indicates the similarity of this image to a real heightmap image.

\par

\begin{figure*}
\centering
\captionsetup[subfigure]{justification=justified,singlelinecheck=false}
\includegraphics[clip,trim=0.06cm 8cm 35.0cm 0.1cm, scale=0.6]{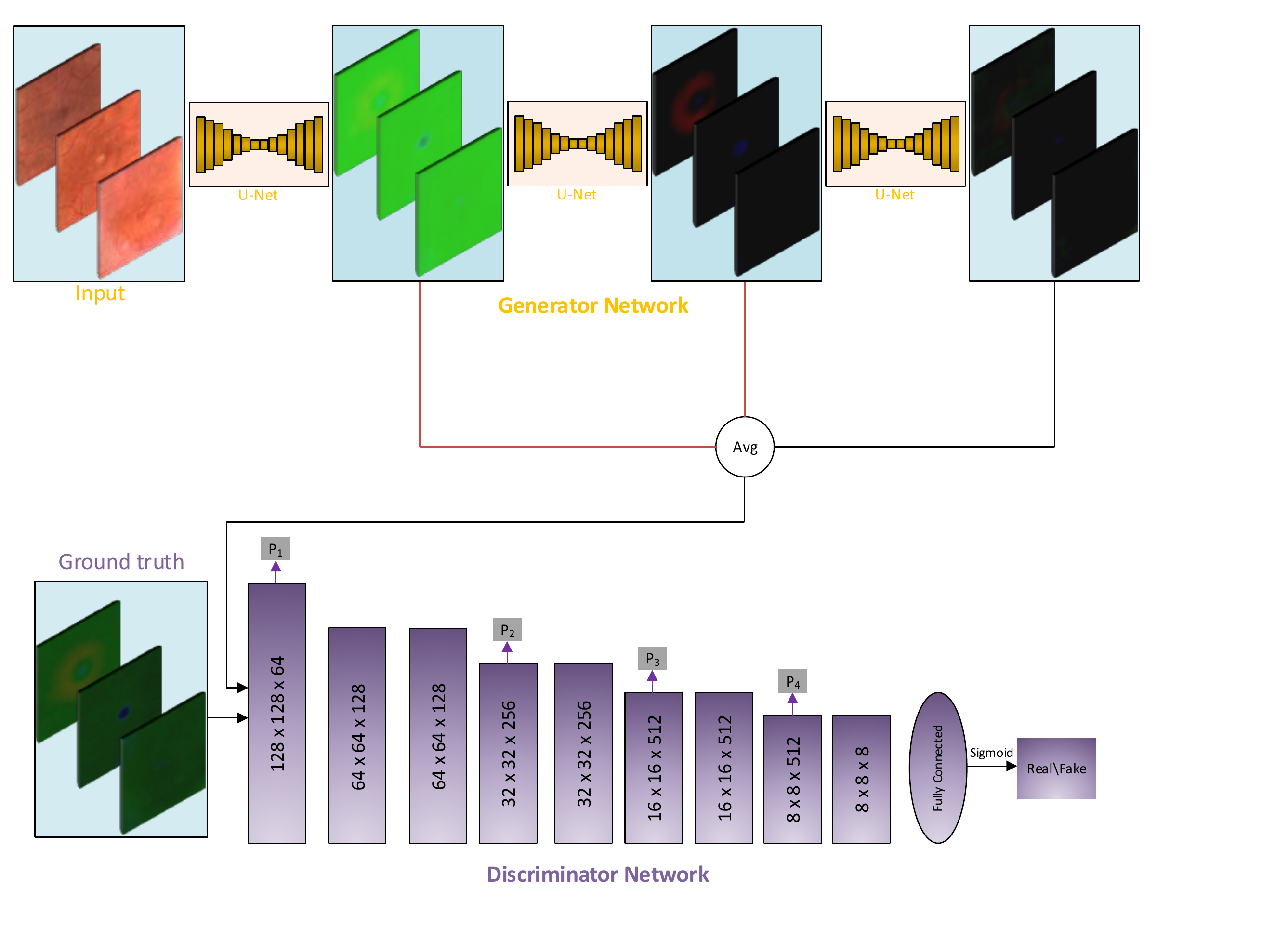}
\caption{The architecture of generator and discriminator in our proposed method. As can be seen, the generator is consists of a series of connected U-Net which we use the output of them for deep supervision (red arrows). Moreover, The discriminator network consists of Convolution-BatchNorm-LeakyReLU layers with a final fully connected layer. We utilized four convolutional layers to compute perceptual loss. The output of the network is in the form of probability and is used to distinguish between real and fake images. }
\label{fig2}
\end{figure*}

Our proposed generator architecture is consists of three stacked U-Nets. Motivated by deeply supervised nets \cite{69} which states that discriminative features in deep layers will contribute to better performance, we used the output of the first two U-Net layers for deep supervision. In fact, similar to U-Net++ architecture \cite{21} which uses the output of the upsampling layer for deep supervision, we averaged the output of all U-Net layers and used the result as the final output of the generator network. By doing so, our network tries to learn a meaningful representation for these deep layers which will directly contribute to the final outcome. Moreover, since the amount of detail is of significance in this work, we decided to use the average operator instead of the max operator for the final output \cite{21}. Another advantage of this architecture is that by using a stack of U-Nets, each layer can add its own level of detail to the final outcome. Furthermore, although our network is deeper in comparison to a normal U-Net architecture, owing to skip-connections and deep supervision involved in the architecture, vanishing gradient problem will not happen and loss flows easily to upper layers through backpropagation. The generator architecture is depicted in the first row of Figure \ref{fig3}.
\par
Regarding the discriminator, the judgment can be made at the image level as well as the patch level. That is to say, we can judge the quality of the entire image by our discriminator (ImageGAN) or consider its patches when we want to judge (PatchGAN). Since a powerful discriminator is the key to successful training with GANs and extremely influences its output \cite{22,27}, we explored both of these methods and opted for image-level discriminator due to better quality images. Furthermore, As can be seen in the second row of Figure \ref{fig3}, we used $1^{th},4^{th},6^{th} $ and $8^{th}$ layer of the discriminator network to compute perceptual loss \cite{40} between generated image and ground-truth image as a supervisory signal with the aim of better output.
\subsection{Objective functions}
Our final loss function is composed of three parts which will be discussed in this section.
\subsubsection{Least-squares adversarial loss}
cGANs are generative models that learn mapping from observed image $x$ and random noise vector $z$ to $\hat{y}$, using generator $G: {x,z} \rightarrow \hat{y}$. Then the discrimnator $D$ aims to classify the concatenation of the source image $x$ and its corresponding ground-truth image $y$ as real $D(x,y) = 1$, while classifying $x$ and the transformed image $\hat{y}$ as fake, $D(x,\hat{y}) = 0$.
\par
Despite performing great in many tasks, GANs suffer from different problems such as mode collapse or unstable training procedure \cite{22}. Therefore, in this work to avoid such problems we adopted Least Square Generative Adversarial Networks (LSGANs) \cite{65}. The idea of LSGAN is that even samples on the right side of the decision boundary can provide signals for training. Hence, for achieving this aim, we adopted the least-squares loss function instead of the traditional cross-entropy loss used in normal GAN to penalize data on the right side of the decision boundary but very far from it. Using this simple yet effective idea we can provide gradients even for samples that are correctly classified by the discriminator. The loss function of LSGAN for both discriminator and generator can be written as follows:

\begin{flalign}
&\underset{D}{\min}\ L_{LSGAN}(D) = \frac{1}{2} \mathbb{E}_{x,y} \Big[\big(D(x,y) - b\big) ^2\Big] \ + \nonumber \\
&\qquad \qquad \qquad \nonumber \frac{1}{2} \mathbb{E}_{x,z}\Big[\Big(D\big(x,G(x,z) \big)-a\Big)^2\Big] \nonumber \\
&\underset{G}{\min}\ L_{LSGAN}(G) = \frac{1}{2} \mathbb{E}_{x,z}\Big[\Big(D\big(x,G(x,z) \big)-c\Big)^2\Big]&&
\end{flalign}

This loss functions directly operates on the logits of the output, where $a = 0$ and $c = b= 1$.
\subsubsection{Pixel reconstruction loss}
Image-to-image translation tasks that rely solely on the adversarial loss function do not produce consistent results \cite{36}. Therefore, we also used pixel reconstruction loss here but we opted for L2-loss rather than widely used L1-loss since it performed better in reconstructing details in this specific task. The equation for L2-loss is as below:
\begin{equation}
\begin{split}
L_{L2}(G) = \mathbb{E}_{x,y,z}\big[\parallel y-G(x,z)\parallel^{2}_{2}\big]
\label{eq3}
\end{split}
\end{equation}
\subsubsection{Perceptual loss}
Despite producing plausible results using only two aforementioned loss functions, since the generated image is blurry \cite{36} and especially in medical diagnosis small details are of significance, we used perceptual loss \cite{40} to improve the final result. As a matter of fact, using only L2-Loss or L1-Loss results in outputs that maintain the global structure but it shows blurriness and distortions \cite{40}. Furthermore, per-pixel losses fail to capture perceptual differences between input and ground-truth images. For instance, when we consider two identical images only shifted by some small offset from each other, per-pixel loss value may vary considerably between these two, even though they are quite similar \cite{39}. However, by using high-level features extracted from layers of a discriminator, we can capture those discrepancies and can measure image similarity more robustly \cite{39}. In our work, since discriminator network also has this capability of perceiving the content of images and difference between them and pre-trained networks on other tasks may perform poorly on other unrelated tasks, we used hidden layers of discriminator network \cite{40,36} to extract features as illustrated in the second row of the Figure \ref{fig2}. The mean absolute error for $i^{th}$ hidden layer between the generated image and the ground-truth image is then calculated as :
\begin{equation}
\begin{split}
P_i\big(G(x,z),y\big) = \frac{1}{w_i h_i d_i} \parallel D_i\big(x,y\big) - D_i\big(x,G(x,z)\big) \parallel _{1}
\label{eq4}
\end{split}
\end{equation}
which $w_i$,$h_i$ and $d_i$ denote width, height and depth of the $i^{th}$ hidden layer respectively and $D_i$ means the output of $i^{th}$ layer of the discriminator network. Finally, perceptual loss can be formulated as :
\begin{equation}
\begin{split}
L_{perceptual} = \sum_{i=0}^{L}\lambda_i P_i\big(G(x,z),y\big)
\label{eq5}
\end{split}
\end{equation}
Where $\lambda_i$ in equation \ref{eq5} tunes the contribution of $i^{th}$ utilized hidden layer on the final loss.
\par
Finally, our complete loss function for the generator is as below:
\begin{equation}
\begin{split}
L = \alpha_{1}L_{perceptual} + \alpha_{2}L_{L2} + \alpha_{3}L_{LSGAN}
\label{eq6}
\end{split}
\end{equation}
where $\alpha_1$, $\alpha_2$ and $\alpha_3$ are the hyperparameters that balance the contribution of each of the different losses. Note that we also used perceptual loss in training discriminator besides traditional cGAN loss with equal contribution.

\section{Experiments}
\subsection{Dataset}
\begin{figure}[h]
\centering
\subfloat{\includegraphics[width=0.24\linewidth]{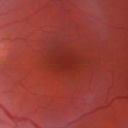}}\
\subfloat{\includegraphics[width=0.24\linewidth]{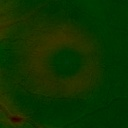}}\
\subfloat{\includegraphics[width=0.24\linewidth]{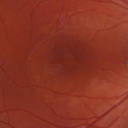}}\
\subfloat{\includegraphics[width=0.24\linewidth]{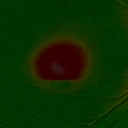}}\
\vfill
\subfloat{\includegraphics[width=0.24\linewidth]{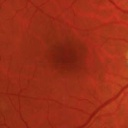}}\
\subfloat{\includegraphics[width=0.24\linewidth]{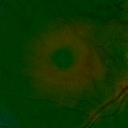}}\
\subfloat{\includegraphics[width=0.24\linewidth]{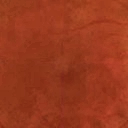}}\
\subfloat{\includegraphics[width=0.24\linewidth]{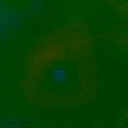}}\

\caption{Macula area on fundus image and their corresponding heightmap in our dataset.}
\label{fig4}
\end{figure}

The dataset was gathered from TopCon DRI OCT Triton captured at Negah Eye Hospital. We cropped the macula part of the fundus and heightmap image from the 3D macula report generated by the device to create image pairs. Our dataset contains 3407 color fundus-heightmap pair images. Since the images in our dataset were insufficient, we used data augmentation for better generalization. Nevertheless, because we are dealing with medical images, we could not rotate images since by rotating images for example by 90$^\circ$, we have vessels in a vertical position which is impossible in fundus imaging. Moreover, changing brightness is also illegal since it will change the standard brightness of a fundus image. Hence, the only augmentation that we could apply was to flip images in 3 different ways to generate 4 different samples from every image. Consequently, we had 13,628 images which we used 80$\%$ for training, 10$\%$ for validation and 10$\%$ for testing. Some examples from the utilized dataset are illustrated in Figure \ref{fig4}.

\subsection{Exprimental setup}
We used Tensorflow 2.0 \cite{66} for implementing our network. We also used Adam optimizer \cite{73} with an initial learning rate of $1e^{-3}$ with a step decay of $0.9$ per $30$ steps. Moreover, we used the batch size of 8 and trained for 250 epochs to converge. Additionally, we set $\lambda_1= 5.0$, $\lambda_2 = 1.0$, $\lambda_3 = 5.0$ and $\lambda_4 = 5.0$ in Equation \ref{eq5} by trial-and-error and considering the contribution of each of them as discussed in \cite{40,39}. Since we are working in a very high dimensional parameter space
convergence and finding the optimal weights is difficult and starting from a random point will not work very well \cite{58}. As a result, in order to ease the training procedure and convergence, we employed a step-by-step training schema. That is to say, we first trained the first U-Net completely, then we added the next U-Net and trained a stack of 2 U-Net with deep supervision and finally, we trained the entire network. Note that we loaded the weights of the previous network when we wanted to train the new one.
\subsection{Evaluation metrics}
In this work, we utilized a variety of different metrics to evaluate our final outcomes quantitatively. We measured the quality of the final image using the Structural Similarity Index (SSIM) \cite{67}, Mean Squared Error (MSE) and Peak Signal to Noise Ratio (PSNR). Nevertheless, these measures are insufficient for assessing structured outputs such as images, as they assume pixel-wise independence \cite{68}. Consequently, we used Learned Perceptual Image Patch Similarity (LPIPS) \cite{68} which can outperform other measures in terms of comparing the perceptual quality of images. We used features extracted from the $1^{th},4^{th},6^{th} $ and $8^{th}$ layer of the discriminator network to obtain the features and calculated the difference between $y$ as the generated heightmap and $ \hat{y}$ as the ground-truth heightmap for given input $x$ using the equation below which all parameters are similar to Equation \ref{eq4}:
\begin{equation}
\begin{split}
d(y,\hat{y},x) = \sum_{l = 0} ^{n}\frac{1}{w_l h_l d_l} \parallel D_l(x,y) - D_l(x,\hat{y})\parallel ^{2}_{2}
\label{eq7}
\end{split}
\end{equation}
\subsection{Analysis of generator architecture}
\begin{figure*}
\subfloat[SSIM and PSNR]{
\begin{minipage}{
0.48\textwidth}
\includegraphics[clip,trim=1.4cm 8.0cm 1cm 1.5cm ,width=1\textwidth]{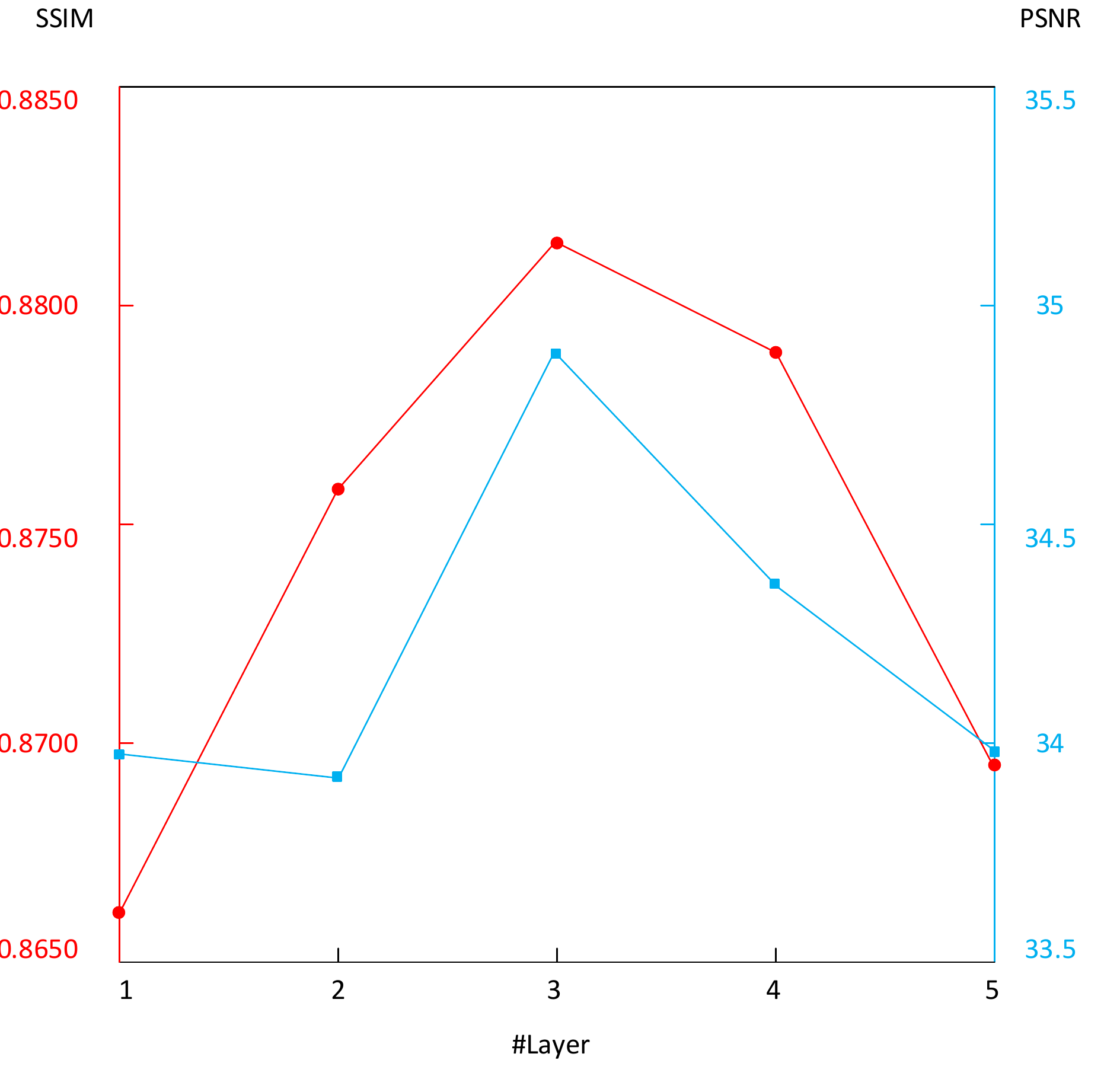}
\end{minipage}}\
\subfloat[MSE and LPIPS]{\begin{minipage}{
0.48\textwidth}
\includegraphics[clip,trim=1.4cm 8.0cm 1cm 1.5cm ,width=1\textwidth]{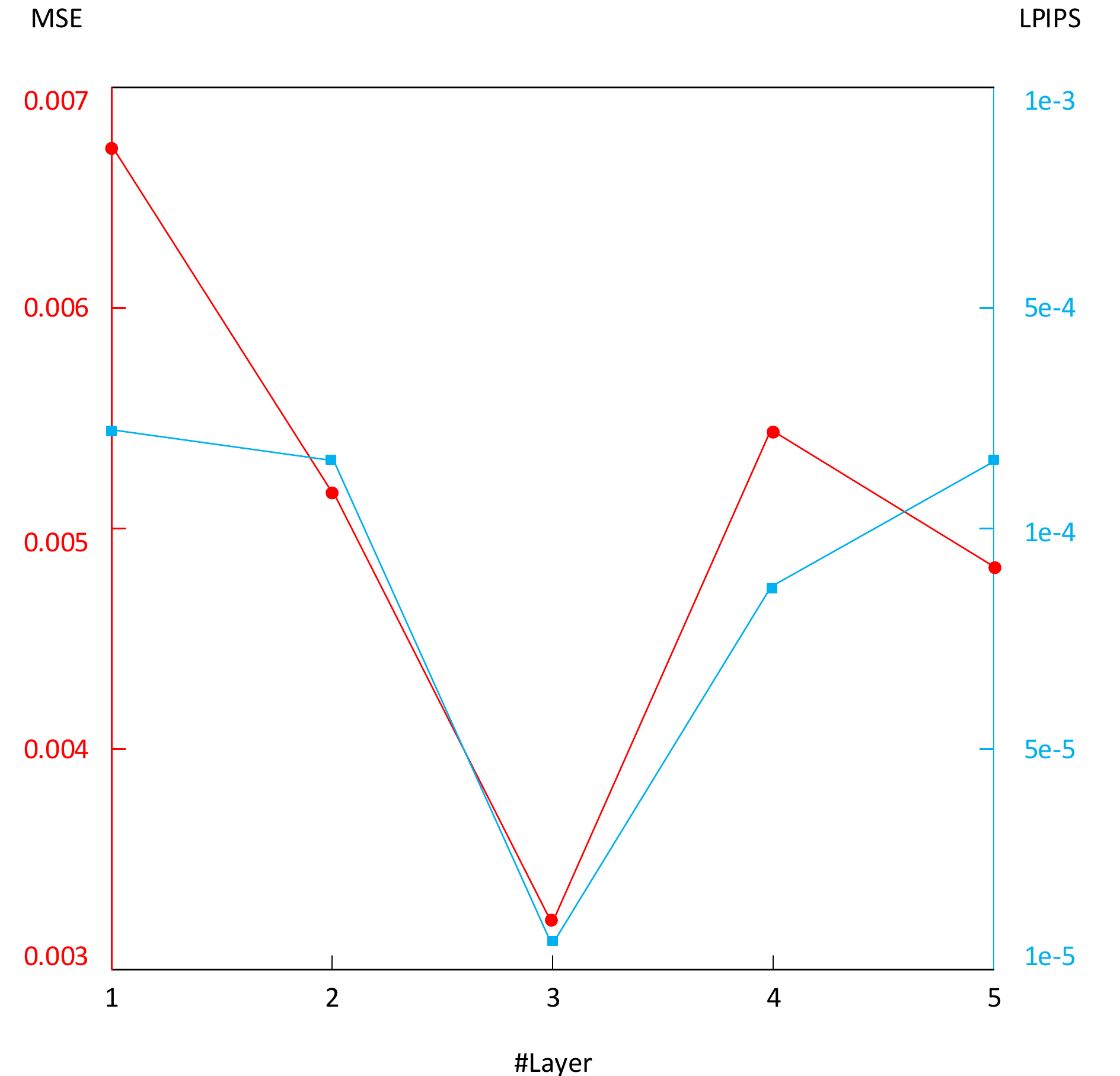}
\end{minipage}}
\caption{Quantitavte comparison between different number of stacked U-Nets in generator architecture.}
\label{fig5}
\end{figure*}

\begin{figure*}

\subfloat{
\begin{minipage}{
0.16\textwidth}
\includegraphics[width=1\textwidth]{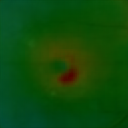}
\end{minipage}}\
\subfloat{\begin{minipage}{
0.16\textwidth}
\includegraphics[width=1\textwidth]{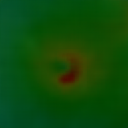}
\end{minipage}}\
\subfloat{\begin{minipage}{
0.16\textwidth}
\includegraphics[width=1\textwidth]{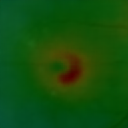}
\end{minipage}}\
\subfloat{\begin{minipage}{
0.16\textwidth}
\includegraphics[width=1\textwidth]{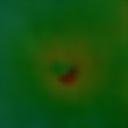}
\end{minipage}}\
\subfloat{\begin{minipage}{
0.16\textwidth}
\includegraphics[width=1\textwidth]{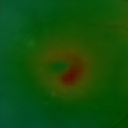}
\end{minipage}}\hspace{0.095cm}
\subfloat{\begin{minipage}{
0.16\textwidth}\hspace{0.095cm}
\includegraphics[width=1\textwidth]{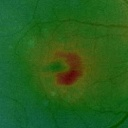}
\end{minipage}}

\setcounter{subfigure}{0}%

\subfloat[1 U-Net]{
\begin{minipage}{
0.16\textwidth}
\includegraphics[width=1\textwidth]{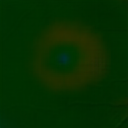}
\end{minipage}}\
\subfloat[2 U-Net]{\begin{minipage}{
0.16\textwidth}
\includegraphics[width=1\textwidth]{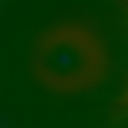}
\end{minipage}}\
\subfloat[3 U-Net]{\begin{minipage}{
0.16\textwidth}
\includegraphics[width=1\textwidth]{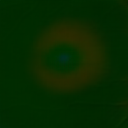}
\end{minipage}}\
\subfloat[4 U-Net]{\begin{minipage}{
0.16\textwidth}
\includegraphics[width=1\textwidth]{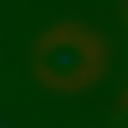}
\end{minipage}}\
\subfloat[5 U-Net]{\begin{minipage}{
0.16\textwidth}
\includegraphics[width=1\textwidth]{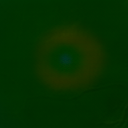}
\end{minipage}}\hspace{0.095cm}
\subfloat[Ground-truth]{\begin{minipage}{
0.16\textwidth}\hspace{0.095cm}
\includegraphics[width=1\textwidth]{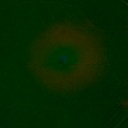}
\end{minipage}}

\caption{Qualitative comparison in terms of different numbers of stacked U-Nets in generator structure.}
\label{fig6}

\end{figure*}

In this part, we explore different numbers of stacked U-Nets in generator architecture to find the optimum one. We set $\alpha_1 = 100$, $\alpha_2 = 1$ and $\alpha_3 = 50$ in Equation \ref{eq6} and used ImageGAN for our discriminator. The quantitative comparison is made in Figure \ref{fig5}. As can be seen, stacking three U-Nets resulted in higher values for SSIM and PSNR and lower values for MSE and LPIPS. Furthermore, qualitative comparison which is depicted in Figure \ref{fig6} also supports our claim that three stacks of U-Nets is the best choice. As a matter of fact, the generator with three U-Nets in this figure did well at predicting the full shape of the red region as well as the correct position and full shape of intense red spots. Additionally, it seems that by adding more U-Nets to the structure, the results become more blurry and details begin to vanish. Therefore, three U-Nets is the optimum number that preserves fine details and can produce plausible outcomes.

\subsection{Effectiveness of deep supervision}

\begin{figure*}
\captionsetup[sub]{labelformat=empty,justification=raggedleft,singlelinecheck=false}
\subfloat{
\begin{minipage}{
0.19\textwidth}
\makebox[0pt][r]{\makebox[20pt]{\raisebox{31pt}{\rotatebox[origin=c]{90}{$w$}}}}%
\includegraphics[width=1\textwidth]{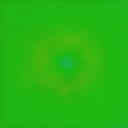}
\end{minipage}}\
\subfloat{\begin{minipage}{
0.19\textwidth}
\includegraphics[width=1\textwidth]{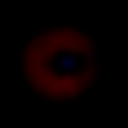}
\end{minipage}}\
\subfloat{\begin{minipage}{
0.19\textwidth}
\includegraphics[width=1\textwidth]{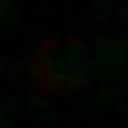}
\end{minipage}}\
\subfloat{\begin{minipage}{
0.19\textwidth}
\includegraphics[width=1\textwidth]{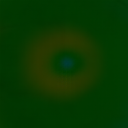}
\end{minipage}}\
\subfloat{\begin{minipage}{
0.19\textwidth}
\includegraphics[width=1\textwidth]{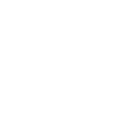}
\end{minipage}}\
\vfill
\vspace{-0.3cm}
\subfloat{
\begin{minipage}{
0.19\textwidth}
\makebox[0pt][r]{\makebox[20pt]{\raisebox{31pt}{\rotatebox[origin=c]{90}{$w/o$}}}}%
\includegraphics[width=1\textwidth]{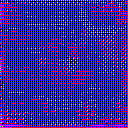}
\end{minipage}}\
\subfloat{\begin{minipage}{
0.19\textwidth}
\includegraphics[width=1\textwidth]{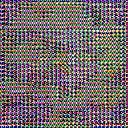}
\end{minipage}}\
\subfloat{\begin{minipage}{
0.19\textwidth}\
\includegraphics[width=1\textwidth]{fig7/blank.png}
\end{minipage}}\
\subfloat{\begin{minipage}{
0.19\textwidth}
\includegraphics[width=1\textwidth]{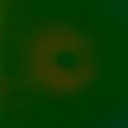}
\end{minipage}}\hspace{0.02cm}
\subfloat{\begin{minipage}{
0.19\textwidth}
\vspace*{-2.6cm}\includegraphics[width=1\textwidth]{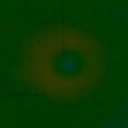}
\end{minipage}}
\vfill
\vspace{-0.2cm}
\subfloat{
\begin{minipage}{
0.19\textwidth}
\makebox[0pt][r]{\makebox[20pt]{\raisebox{31pt}{\rotatebox[origin=c]{90}{$w$}}}}%
\includegraphics[width=1\textwidth]{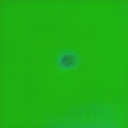}
\end{minipage}}\
\subfloat{\begin{minipage}{
0.19\textwidth}
\includegraphics[width=1\textwidth]{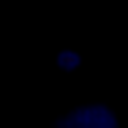}
\end{minipage}}\
\subfloat{\begin{minipage}{
0.19\textwidth}
\includegraphics[width=1\textwidth]{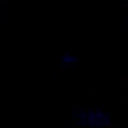}
\end{minipage}}\
\subfloat{\begin{minipage}{
0.19\textwidth}
\includegraphics[width=1\textwidth]{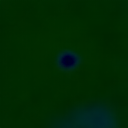}
\end{minipage}}\
\subfloat{\begin{minipage}{
0.19\textwidth}
\includegraphics[width=1\textwidth]{fig7/blank.png}
\end{minipage}}\
\vfill
\vspace{-0.3cm}
\subfloat{
\begin{minipage}{
0.19\textwidth}
\makebox[0pt][r]{\makebox[20pt]{\raisebox{31pt}{\rotatebox[origin=c]{90}{$w/o$}}}}%
\includegraphics[width=1\textwidth]{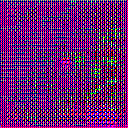}
\end{minipage}}\
\subfloat{\begin{minipage}{
0.19\textwidth}
\includegraphics[width=1\textwidth]{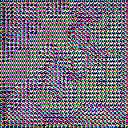}
\end{minipage}}\
\subfloat{\begin{minipage}{
0.19\textwidth}\
\includegraphics[width=1\textwidth]{fig7/blank.png}
\end{minipage}}\
\subfloat{\begin{minipage}{
0.19\textwidth}
\includegraphics[width=1\textwidth]{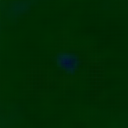}
\end{minipage}}\hspace{0.02cm}
\subfloat{\begin{minipage}{
0.19\textwidth}
\vspace*{-2.6cm}\includegraphics[width=1\textwidth]{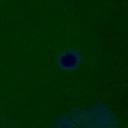}
\end{minipage}}
\vfill
\vspace{-0.2cm}
\subfloat{
\begin{minipage}{
0.19\textwidth}
\makebox[0pt][r]{\makebox[20pt]{\raisebox{31pt}{\rotatebox[origin=c]{90}{$w$}}}}%
\includegraphics[width=1\textwidth]{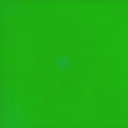}
\end{minipage}}\
\subfloat{\begin{minipage}{
0.19\textwidth}
\includegraphics[width=1\textwidth]{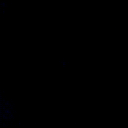}
\end{minipage}}\
\subfloat{\begin{minipage}{
0.19\textwidth}
\includegraphics[width=1\textwidth]{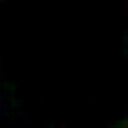}
\end{minipage}}\
\subfloat{\begin{minipage}{
0.19\textwidth}
\includegraphics[width=1\textwidth]{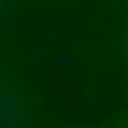}
\end{minipage}}\
\subfloat{\begin{minipage}{
0.19\textwidth}
\includegraphics[width=1\textwidth]{fig7/blank.png}
\end{minipage}}\
\vfill
\vspace{-0.3cm}
\setcounter{subfigure}{0}%

\subfloat[First layer]{
\begin{minipage}{
0.19\textwidth}
\makebox[0pt][r]{\makebox[20pt]{\raisebox{31pt}{\rotatebox[origin=c]{90}{$w/o$}}}}%
\includegraphics[width=1\textwidth]{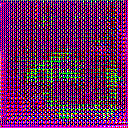}
\end{minipage}}\
\subfloat[Second layer]{\begin{minipage}{
0.19\textwidth}
\includegraphics[width=1\textwidth]{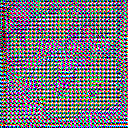}
\end{minipage}}\
\subfloat[Third layer]{\begin{minipage}{
0.19\textwidth}\
\includegraphics[width=1\textwidth]{fig7/blank.png}
\end{minipage}}\
\subfloat[Output]{\begin{minipage}{
0.19\textwidth}
\includegraphics[width=1\textwidth]{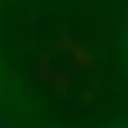}
\end{minipage}}\hspace{0.02cm}
\captionsetup[subfigure]{captionskip=1.06cm,margin={-0.7cm,-0.1cm}}
\subfloat[Ground-truth]{\begin{minipage}{
0.19\textwidth}
\vspace*{-2.6cm}\includegraphics[width=1\textwidth]{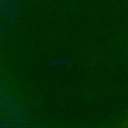}
\end{minipage}}

\caption{The effect of deep supervision on the final output. As can be seen, the outputs from the first and second layer of the model with deep supervision(w) generated meaningful results and represent the parts on which the layer focused. On the other hand, without the use of deep supervision(w/o), generated images from deep layers do not contain useful information and caused lower output quality.}
\label{fig7}
\end{figure*}

In this section, we explore the effectiveness of using deep supervision. For this experiment, we trained our network with and without the use of deep supervision. As can be seen in Figure \ref{fig7}, the output of the deep layers when we used deep supervision is meaningful. In fact, by averaging the output of each of the layers, we force the network to output plausible images from these layers and this contributes to the higher quality of the network with deep supervision.
\par
As depicted in Figure \ref{fig7}, the output of the first U-Net layer is responsible for the overall brightness of the output image besides vaguely representing blue and red parts. The output of the second U-Net is mostly used for abnormal regions that are overlayed on the green parts. In fact, it is responsible for detecting higher or lower elevated parts in the macula region. Finally, the third layer is responsible for adding fine details to the output. If the image given does not contain any abnormalities, the output from the second and third deep layer is mostly black (e.g. Figure \ref{fig7} third example). However, considering the model without supervision, clearly, there is not any meaningful interpretation for the images outputted from the first and second layer and this contributed to the lower overall quality of this model. Quantitative comparisons in Table \ref{tb1} also proves our point that deep supervision can contribute to the finer output. As can be seen, our model with deep supervision achieved a higher score in all evaluation metrics.

\begin{table}[!ht]
\centering
\begin{tabular}{l|l|l|l|l}
& SSIM & LPIPS & MSE & PSNR(dB) \\ \hline
w supervision & \textbf{0.8823} & \textbf{1.81e-05} & \textbf{0.0033} & \textbf{34.6733} \\ \hline
w\textbackslash{}o supervision & 0.7570 & 2.54e-04 & 0.0059 & 27.2784
\end{tabular}
\caption{Quantitative comparison between model trained with supervision and model trained without supervision.}
\label{tb1}
\end{table}

\subsection{L1-Loss vs L2-Loss}

Even though in most of the papers L1-Loss is more common than L2-Loss as pixel-loss reconstruction loss \cite{75,24,36}, in this work we chose L2-Loss owing to emphasis that L2-Loss put on huge differences between generated image and ground-truth. As a matter of fact, since the difference in L2-Loss has a power of two, small differences become minuscule and negligible and the focus will be on huge differences. This behavior is perfectly suitable for this problem since our important goal is to predict regions that have a red color or blue color and it is acceptable to have inaccurate or blurry green areas or missed vessels. This is because those red or blue regions contain significant information for diagnosis since they are related to regions with elevation changes which are important in the diagnosis of many retinal diseases such as PED which cause different parts of the macula to swell. Our claim is supported by our experiment in which we compared the results from L1-Loss and L2-Loss. Note that in this experiment the contribution of L2-Loss and L1-Loss function was equal along with LSGAN and perceptual loss. As can be seen in Table \ref{tb2}, L2-Loss performed better in all metrics and the difference is considerable.

\begin{table}[!ht]
\centering
\begin{tabular}{l|l|l|l|l}
& SSIM & LPIPS & MSE & PSNR(dB) \\ \hline
L1-Loss & 0.8721 & 3.53e-04 & 0.0072 & 33.8351 \\ \hline
L2-Loss & \textbf{0.8823} & \textbf{1.81e-05} & \textbf{0.0033} & \textbf{34.6733}
\end{tabular}
\caption{Quantitative comparison between L2-Loss and L1-Loss.}
\label{tb2}
\end{table}

\begin{figure}[h]
\subfloat{
\begin{minipage}{
0.31\linewidth}
\includegraphics[width=1\textwidth]{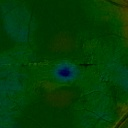}
\end{minipage}}\hspace{0.05cm}
\subfloat{\begin{minipage}{
0.31\linewidth}
\includegraphics[width=1\textwidth]{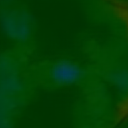}
\end{minipage}}\
\subfloat{\begin{minipage}{
0.31\linewidth}
\includegraphics[width=1\textwidth]{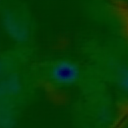}
\end{minipage}}

\setcounter{subfigure}{0}%

\subfloat[Ground-truth]{
\begin{minipage}{
0.31\linewidth}
\includegraphics[width=1\textwidth]{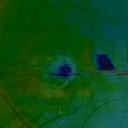}
\end{minipage}}\hspace{0.05cm}
\subfloat[L1]{\begin{minipage}{
0.31\linewidth}
\includegraphics[width=1\textwidth]{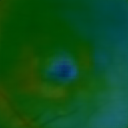}
\end{minipage}}\
\subfloat[L2]{\begin{minipage}{
0.31\linewidth}
\includegraphics[width=1\textwidth]{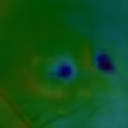}
\end{minipage}}

\caption{Qualitative comparison between the output of L1-Loss and L2-Loss.}
\label{fig8}

\end{figure}

Furthermore, regarding qualitative comparison in Figure \ref{fig8}, even though the global structure of images considering green areas is roughly the same, L2-Loss performed better at predicting blue regions which is crucial for diagnosis.

\subsection{Comparison with other techniques}
\label{sec5.8}
\begin{figure*}
\centering

\subfloat{
\begin{minipage}{
0.19\textwidth}
\includegraphics[width=1\textwidth,height = 1\textwidth]{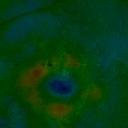}
\end{minipage}}\hspace{0.25cm}
\subfloat{\begin{minipage}{
0.19\textwidth}
\includegraphics[width=1\textwidth,height = 1\textwidth]{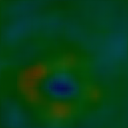}
\end{minipage}}\
\subfloat{\begin{minipage}{
0.19\textwidth}
\includegraphics[width=1\textwidth,height = 1\textwidth]{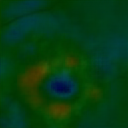}
\end{minipage}}\
\subfloat{\begin{minipage}{
0.19\textwidth}
\includegraphics[width=1\textwidth,height = 1\textwidth]{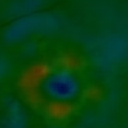}
\end{minipage}}\
\subfloat{\begin{minipage}{
0.19\textwidth}
\includegraphics[width=1\textwidth,height = 1\textwidth]{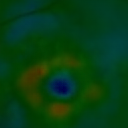}
\end{minipage}}

\subfloat{
\begin{minipage}{
0.19\textwidth}
\includegraphics[width=1\textwidth,height = 1\textwidth]{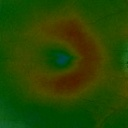}
\end{minipage}}\hspace{0.25cm}
\subfloat{\begin{minipage}{
0.19\textwidth}
\includegraphics[width=1\textwidth,height = 1\textwidth]{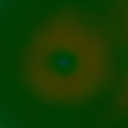}
\end{minipage}}\
\subfloat{\begin{minipage}{
0.19\textwidth}
\includegraphics[width=1\textwidth,height = 1\textwidth]{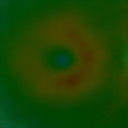}
\end{minipage}}\
\subfloat{\begin{minipage}{
0.19\textwidth}
\includegraphics[width=1\textwidth,height = 1\textwidth]{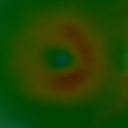}
\end{minipage}}\
\subfloat{\begin{minipage}{
0.19\textwidth}
\includegraphics[width=1\textwidth,height = 1\textwidth]{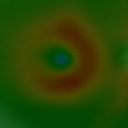}
\end{minipage}}

\setcounter{subfigure}{0}%

\subfloat[Ground-truth]{
\begin{minipage}{
0.19\textwidth}
\includegraphics[width=1\textwidth,height = 1\textwidth]{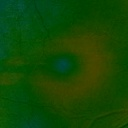}
\end{minipage}}\hspace{0.25cm}
\subfloat[pix2pix]{\begin{minipage}{
0.19\textwidth}
\includegraphics[width=1\textwidth,height = 1\textwidth]{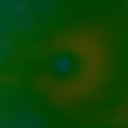}
\end{minipage}}\
\subfloat[PAN]{\begin{minipage}{
0.19\textwidth}
\includegraphics[width=1\textwidth,height = 1\textwidth]{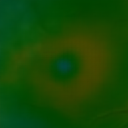}
\end{minipage}}\
\subfloat[MedGAN]{\begin{minipage}{
0.19\textwidth}
\includegraphics[width=1\textwidth,height = 1\textwidth]{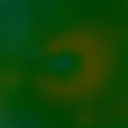}
\end{minipage}}\
\subfloat[Proposed method]{\begin{minipage}{
0.19\textwidth}
\includegraphics[width=1\textwidth,height = 1\textwidth]{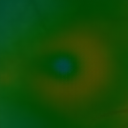}
\end{minipage}}
\setcounter{subfigure}{0}%

\caption{Qualitative comparison between the proposed method and other methods.}
\label{fig9}

\end{figure*}

Since this is the first method for the reconstruction of the heightmap of color fundus images using DNNs, there were no other method to directly compare our proposed method with. Therefore, we compared the results with popular methods that utilized cGANs. The methods that we compared our results to here are pix2pix \cite{24}, PAN \cite{40} and MedGAN \cite{36}. The results are given in Figure \ref{fig9} and Table \ref{tb3} for qualitative and quantitative comparison respectively. Pix2pix achieved the worst results since it does not use deep supervision and it is based on L1-loss. PAN did slightly better in SSIM and MSE metrics. However, there is a huge difference between PAN and pix2pix in terms of LPIPS since PAN uses perceptual loss in the training procedure. This difference proves the importance and the impact of using perceptual loss for training cGANs. MedGAN was designed especially for medical image translation such as translating between CT and PET images. As a result, It performs better in comparison to previous general methods, but the results are inferior to our proposed method. In fact, since we are using deep supervision in this method and carefully tuned the parameters for this particular problem, we achieve a higher value in all metrics. Another justification for higher values of the proposed method is that all the previous methods used L1-loss as pixel-reconstruction loss for the training, while we used L2-loss which as stated in the previous section, is the superior choice for this particular problem.

\begin{table*}[!ht]
\centering
\begin{tabular}{l|c|c|c|c}
Method & \multicolumn{1}{l|}{SSIM} & \multicolumn{1}{l|}{PSNR(dB)} & \multicolumn{1}{l|}{LPIPS} & \multicolumn{1}{l}{MSE} \\ \hline
pix2pix \cite{24} & 0.8596 & 33.9523 & 2.25e-03 & 0.0068 \\ \cline{1-1}
PAN \cite{40} & 0.8612 & 33.8512 & 2.37e-04 & 0.0053 \\ \cline{1-1}
MedGAN \cite{36} & 0.8659 & 33.2958 & 5.61e-05 & 0.0048 \\ \cline{1-1}
Proposed Method & \textbf{0.8823} &\textbf{ 34.6733} & \textbf{1.81e-05} &\textbf{0.0033}
\end{tabular}
\caption{Quantitative comparison between proposed method and other methods.}
\label{tb3}
\end{table*}

\par

Considering the qualitative comparison in Figure \ref{fig9}, our proposed method outperformed others in terms of reconstruction of the details. As can be seen, pix2pix missed some of the important details in the first and second examples such as bright red spots. PAN performed better at reconstructing the highly elevated parts in the second row, however it failed to reconstruct the correct shape for the third example. Finally, since MedGAN is specially designed for medical tasks, it outperformed the aforementioned methods in terms of output quality, but it was outperformed by our method and the proposed method generated the best quality images.

\subsection{Perceptual studies}
\label{sectionPerceptualStudies}
As stated before, the main purpose of reconstructing the heightmap of fundus images is to be able to infer part of the information that OCT devices provide to ophthalmologists, namely the height information. Hence, to judge the fidelity of the reconstructed heightmap, in this section we conducted an experiment in which two experienced ophthalmologists were presented a series of trails each containing reconstructed heightmap, fundus image and the ground-truth heightmap from the test set. The main purpose of this study is to investigate if the reconstructed heightmap and fundus image pair gives more information for the diagnosis of any retinal disease in comparison to the situation in which we have the fundus image only.

\par

For this experiment, two ophthalmologists first classified all images into two classes positive and negative which positive means that the image provides more information for diagnosis and negative means that the image does not add more information for diagnosis. Additionally, ophthalmologists rated each image from zero to three according to the level of information that each of the given images provide such that zero means no added information and three represents the highest amount of information for diagnosis.
\par

As can be seen in Table \ref{tb4}, ophthalmologist 1 classified all images as useful for diagnosis and the mean score for all of the images is 1.94. Additionally, ophthalmologist 2 classified 92$\%$ of the outputs as positive and the mean square is 1.84. This study shows that even though the output of our method may seem more blurry in comparison to the original one, these outputs can be used for diagnosis and can provide valuable additional information to ophthalmologists especially about height information in different regions. For instance, diseases such as Age-related macular degeneration are dependent on the swelling of different regions of the macula and the reconstructed heightmap contains this information.

\par

We also considered the positive samples in isolation and the results are shown in Table \ref{tb5}. As can be seen, both ophthalmologists classified most of the images in class 2 and the average score is near 2 for both ophthalmologists. This experiment also indicates that in most of the cases the reconstructed heightmap can provide useful information for diagnosis from a single fundus image.

\begin{table}[h]
\centering
\begin{tabular}{l|c|c|c}
\multirow{2}{*}{} & \multicolumn{2}{c|}{Score} & Classification \\ \cline{2-4}
& Mean & SD & Positive \% \\ \hline
Ophthalmologist 1 & 1.94 & 0.7669 & 100.00 \\ \cline{1-1}
Ophthalmologist 2 & 1.84 & 0.9553 & 92.00
\end{tabular}
\caption{Results of perceptual study.}
\label{tb4}
\end{table}

\begin{table*}[]
\begin{tabular}{c|c|c|c|c|cccc}
& \multicolumn{4}{c|}{Ophthalmologist 1} & \multicolumn{4}{c}{Ophthalmologist 2} \\ \hline
Score & Frequency & \% & Mean & SD & \multicolumn{1}{c|}{Frequency} & \multicolumn{1}{c|}{\%} & \multicolumn{1}{c|}{Mean} & SD \\ \hline
1 & 16 & 32.00 & \multirow{3}{*}{1.94} & & \multicolumn{1}{c|}{15} & \multicolumn{1}{c|}{30.00} & \multicolumn{1}{c|}{\multirow{3}{*}{2.00}} & \multirow{3}{*}{0.8164} \\
2 & 21 & 42.00 & & 0.7669 & \multicolumn{1}{c|}{16} & \multicolumn{1}{c|}{32.00} & \multicolumn{1}{c|}{} & \\
3 & 13 & 26.00 & & & \multicolumn{1}{c|}{15} & \multicolumn{1}{c|}{30.00} & \multicolumn{1}{c|}{} &
\end{tabular}
\caption{Information of positive samples in perceptual studies.}
\label{tb5}
\end{table*}

\begin{figure*}
\captionsetup[subfigure]{labelformat=empty,position=top}
\centering

\subfloat[Fundus]{
\begin{minipage}{
0.15\textwidth}
\includegraphics[width=1\textwidth,height = 1\textwidth]{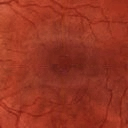}
\end{minipage}}\
\subfloat[Proposed method]{\begin{minipage}{
0.15\textwidth}
\includegraphics[width=1\textwidth,height = 1\textwidth]{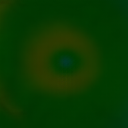}
\end{minipage}}\
\subfloat[Ground-truth]{\begin{minipage}{
0.15\textwidth}
\includegraphics[width=1\textwidth,height = 1\textwidth]{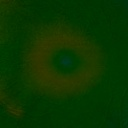}
\end{minipage}}\hspace{0.25cm}
\subfloat[Fundus]{\begin{minipage}{
0.15\textwidth}
\includegraphics[width=1\textwidth,height = 1\textwidth]{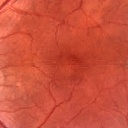}
\end{minipage}}\
\subfloat[Proposed method]{\begin{minipage}{
0.15\textwidth}
\includegraphics[width=1\textwidth,height = 1\textwidth]{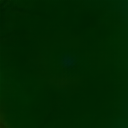}
\end{minipage}}\
\subfloat[Ground-truth]{\begin{minipage}{
0.15\textwidth}
\includegraphics[width=1\textwidth,height = 1\textwidth]{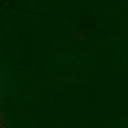}
\end{minipage}}

\captionsetup[subfigure]{labelformat=parens,position=bottom}
\setcounter{subfigure}{0}%
\subfloat{
\begin{minipage}{
0.15\textwidth}
\includegraphics[width=1\textwidth,height = 1\textwidth]{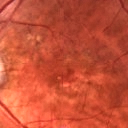}
\end{minipage}}\
\captionsetup[subfigure]{captionskip=0.5cm,margin={-0.1cm,0.0cm}}
\setcounter{subfigure}{0}%
\subfloat{\begin{minipage}{
0.15\textwidth}
\includegraphics[width=1\textwidth,height = 1\textwidth]{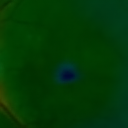}
\end{minipage}}\
\subfloat{\begin{minipage}{
0.15\textwidth}
\includegraphics[width=1\textwidth,height = 1\textwidth]{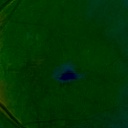}
\end{minipage}}\hspace{0.25cm}
\subfloat{\begin{minipage}{
0.15\textwidth}
\includegraphics[width=1\textwidth,height = 1\textwidth]{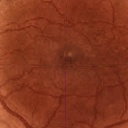}
\end{minipage}}\
\setcounter{subfigure}{1}%
\subfloat{\begin{minipage}{
0.15\textwidth}
\includegraphics[width=1\textwidth,height = 1\textwidth]{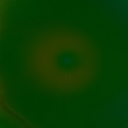}
\end{minipage}}\
\subfloat{\begin{minipage}{
0.15\textwidth}
\includegraphics[width=1\textwidth,height = 1\textwidth]{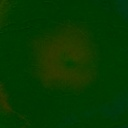}
\end{minipage}}

\caption{Some examples from images classified as positive by two ophthalmologists.}
\label{fig10}

\end{figure*}

Considering examples which were classified as positive in Figure \ref{fig10}, the reconstructed heightmap can indicate the lack of elevation change (top right example) as well as serious elevation changes in different regions (other examples) depending on the condition of the fundus image. In fact, both lack of elevation changes and having high or low elevated parts are types of information that cannot be inferred solely from a single fundus image. However, by utilizing our proposed method, ophthalmologists can be exposed to this additional information which can aid them in better and easier diagnosis with a single color fundus image. Finally, this figure also proves the point in the last paragraph that even though the reconstructed heightmaps in positive samples may seem more blurry than the ground-truth heightmaps, they were classified as positive owing to the information that they provide for diagnosis.

\section{Conclusion and Discussion}
In this paper, we proposed a novel framework to automatically generate a heightmap image of the macula on a color fundus image. For the generator network, we used a stack of three U-Nets and motivated by deeply supervised networks, we averaged on the output of these U-Net layers for deep supervision. We also utilized LSGAN instead of traditional GAN loss for stable training procedure and better results along with L2-Loss and perceptual loss to generate the final outcome.
\par
The experimental results indicate that our proposed method outperformed other methods in the task of image and medical image translation in terms of SSIM, PSNR, MSE and LPIPS metrics as can be seen in Table \ref{tb3}. Furthermore, as depicted in Figure \ref{fig7}, deep supervision contributed greatly to the quality of the final outcome by producing meaningful outputs from deep layers. This suggests that when we are dealing with very deep neural networks, it is better to somehow constrain deep layers into generating features toward the final goal of the network. Finally, considering the applications of our proposed method in real diagnosis, perceptual studies in Table \ref{tb4} and \ref{tb5} indicate that we can infer more information from the reconstructed heightmap, especially in cases in which there are some elevation changes in different regions of the macula region. By utilizing this information about heights, we can provide additional information for the diagnosis of diseases that are dependent on the presence of data about elevations using only a color fundus image and without the need for OCT images.
\par
As stated in Section \ref{sectionPerceptualStudies}, despite slight blurriness in the output of our proposed network, it can still be used for diagnosis. However, this work is not free from limitations with further improvements is essential for improving the practical applicability in real diagnosis cases. In fact, in some cases owing to the poor image quality of the fundus image the system cannot extract useful and meaningful information from the image especially in cases in which the fundus image is blurred. This suggests that in future works, a pre-processing step should be employed to de-blur fundus images properly before feeding them into the network and study its effectiveness. Furthermore, in future works, we will try to utilize other features of the fundus image besides automatic features extracted from CNNs to improve the overall performance and quality of the proposed method.
\par
In future works, we will utilize images from different regions of fundus image such as Optic Nerve Head(ONH) to reconstruct its heightmap as this part of the fundus image has many practical applications and to develop our method into a general solution for heightmap reconstruction. Finally, considering the perceptual studies which show that our reconstructed heightmap can provide information for the diagnosis, in future researches, we can utilize the results generated from our network to detect retinal diseases automatically which were impossible before using only a single fundus image.

\section{Acknowledgment}
The authers are grateful to Dr.Ahmadieh (Ophthalmologist) for grading and classifying images for our experiment.

\bibliographystyle{cas-model2-names}

\bibliography{Refrences}

\begin{thebibliography}{35}
\expandafter\ifx\csname natexlab\endcsname\relax\def\natexlab#1{#1}\fi
\providecommand{\url}[1]{\texttt{#1}}
\providecommand{\href}[2]{#2}
\providecommand{\path}[1]{#1}
\providecommand{\DOIprefix}{doi:}
\providecommand{\ArXivprefix}{arXiv:}
\providecommand{\URLprefix}{URL: }
\providecommand{\Pubmedprefix}{pmid:}
\providecommand{\doi}[1]{\href{http://dx.doi.org/#1}{\path{#1}}}
\providecommand{\Pubmed}[1]{\href{pmid:#1}{\path{#1}}}
\providecommand{\bibinfo}[2]{#2}
\ifx\xfnm\relax \def\xfnm[#1]{\unskip,\space#1}\fi
\bibitem[{Abadi et~al.(2016)Abadi, Barham, Chen, Chen, Davis, Dean, Devin,
  Ghemawat, Irving, Isard, Kudlur, Levenberg, Monga, Moore, Murray, Steiner,
  Tucker, Vasudevan, Warden and Zhang}]{66}
\bibinfo{author}{Abadi, M.}, \bibinfo{author}{Barham, P.},
  \bibinfo{author}{Chen, J.}, \bibinfo{author}{Chen, Z.},
  \bibinfo{author}{Davis, A.}, \bibinfo{author}{Dean, J.},
  \bibinfo{author}{Devin, M.}, \bibinfo{author}{Ghemawat, S.},
  \bibinfo{author}{Irving, G.}, \bibinfo{author}{Isard, M.},
  \bibinfo{author}{Kudlur, M.}, \bibinfo{author}{Levenberg, J.},
  \bibinfo{author}{Monga, R.}, \bibinfo{author}{Moore, S.},
  \bibinfo{author}{Murray, D.}, \bibinfo{author}{Steiner, B.},
  \bibinfo{author}{Tucker, P.}, \bibinfo{author}{Vasudevan, V.},
  \bibinfo{author}{Warden, P.}, \bibinfo{author}{Zhang, X.},
  \bibinfo{year}{2016}.
\newblock \bibinfo{title}{Tensorflow: A system for large-scale machine
  learning} .
\bibitem[{Afsun et~al.(2015)Afsun, Maryam, Farhad, Arezoo, Masoumeh, Zahra,
  Nouraddin, Yassin, Maryam, Poorya and Sosan}]{2}
\bibinfo{author}{Afsun, N.}, \bibinfo{author}{Maryam, G.},
  \bibinfo{author}{Farhad, A.}, \bibinfo{author}{Arezoo, N.},
  \bibinfo{author}{Masoumeh, H.S.}, \bibinfo{author}{Zahra, M.},
  \bibinfo{author}{Nouraddin, K.}, \bibinfo{author}{Yassin, L.},
  \bibinfo{author}{Maryam, H.}, \bibinfo{author}{Poorya, Y.},
  \bibinfo{author}{Sosan, G.}, \bibinfo{year}{2015}.
\newblock \bibinfo{title}{Prevalence of eye disorders in elderly population of
  tehran, iran}.
\bibitem[{Armanious et~al.(2019)Armanious, Yang, Fischer, K{\"u}stner,
  Nikolaou, Gatidis and Yang}]{36}
\bibinfo{author}{Armanious, K.}, \bibinfo{author}{Yang, C.},
  \bibinfo{author}{Fischer, M.}, \bibinfo{author}{K{\"u}stner, T.},
  \bibinfo{author}{Nikolaou, K.}, \bibinfo{author}{Gatidis, S.},
  \bibinfo{author}{Yang, B.}, \bibinfo{year}{2019}.
\newblock \bibinfo{title}{Medgan: Medical image translation using gans}.
\newblock \bibinfo{journal}{Computerized medical imaging and graphics : the
  official journal of the Computerized Medical Imaging Society}
  \bibinfo{volume}{79}, \bibinfo{pages}{101684}.
\bibitem[{Ben-Cohen et~al.(2019)Ben-Cohen, Klang, Raskin, Soffer, Ben-Haim,
  Konen, Amitai and Greenspan}]{74}
\bibinfo{author}{Ben-Cohen, A.}, \bibinfo{author}{Klang, E.},
  \bibinfo{author}{Raskin, S.}, \bibinfo{author}{Soffer, S.},
  \bibinfo{author}{Ben-Haim, S.}, \bibinfo{author}{Konen, E.},
  \bibinfo{author}{Amitai, M.}, \bibinfo{author}{Greenspan, H.},
  \bibinfo{year}{2019}.
\newblock \bibinfo{title}{Cross-modality synthesis from ct to pet using fcn and
  gan networks for improved automated lesion detection}.
\newblock \bibinfo{journal}{Eng. Appl. Artif. Intell.} \bibinfo{volume}{78},
  \bibinfo{pages}{186--194}.
\bibitem[{Brown et~al.(2003)Brown, Brown, Sharma and Busbee}]{1}
\bibinfo{author}{Brown, M.}, \bibinfo{author}{Brown, G.},
  \bibinfo{author}{Sharma, S.}, \bibinfo{author}{Busbee, B.},
  \bibinfo{year}{2003}.
\newblock \bibinfo{title}{Quality of life associated with visual loss: a time
  tradeoff utility analysis comparison with medical health states}.
\newblock \bibinfo{journal}{Ophthalmology} \bibinfo{volume}{110},
  \bibinfo{pages}{1076--81}.
\bibitem[{Cheriyan et~al.(2012)Cheriyan, Hema, Menon and K.A.}]{60}
\bibinfo{author}{Cheriyan, J.}, \bibinfo{author}{Hema, B.},
  \bibinfo{author}{Menon, H.}, \bibinfo{author}{K.A., N.},
  \bibinfo{year}{2012}.
\newblock \bibinfo{title}{3d reconstruction of human retina from a single
  fundus image} .
\bibitem[{Ebenezer et~al.(2019)Ebenezer, Das and Mukhopadhyay}]{77}
\bibinfo{author}{Ebenezer, J.P.}, \bibinfo{author}{Das, B.},
  \bibinfo{author}{Mukhopadhyay, S.}, \bibinfo{year}{2019}.
\newblock \bibinfo{title}{Single image haze removal using conditional
  wasserstein generative adversarial networks}.
\newblock \bibinfo{journal}{2019 27th European Signal Processing Conference
  (EUSIPCO)} \URLprefix \url{http://dx.doi.org/10.23919/EUSIPCO.2019.8902992},
  \DOIprefix\doi{10.23919/eusipco.2019.8902992}.
\bibitem[{Goodfellow et~al.(2014)Goodfellow, Pouget-Abadie, Mirza, Xu,
  Warde-Farley, Ozair, Courville and Bengio}]{22}
\bibinfo{author}{Goodfellow, I.J.}, \bibinfo{author}{Pouget-Abadie, J.},
  \bibinfo{author}{Mirza, M.}, \bibinfo{author}{Xu, B.},
  \bibinfo{author}{Warde-Farley, D.}, \bibinfo{author}{Ozair, S.},
  \bibinfo{author}{Courville, A.}, \bibinfo{author}{Bengio, Y.},
  \bibinfo{year}{2014}.
\newblock \bibinfo{title}{Generative adversarial networks}.
\newblock \href{http://arxiv.org/abs/1406.2661}{\tt arXiv:1406.2661}.
\bibitem[{Guo et~al.(2018)Guo, Zhao, Zou and Ouyang}]{62}
\bibinfo{author}{Guo, F.}, \bibinfo{author}{Zhao, X.}, \bibinfo{author}{Zou,
  B.}, \bibinfo{author}{Ouyang, P.}, \bibinfo{year}{2018}.
\newblock \bibinfo{title}{3d reconstruction and registration for retinal image
  pairs}, pp. \bibinfo{pages}{364--368}.
\newblock \DOIprefix\doi{10.1109/ICIVC.2018.8492769}.
\bibitem[{Huang et~al.(1991)Huang, EA, CP, Schuman, WG, W, Hee, T, K, CA and
  JG}]{61}
\bibinfo{author}{Huang, D.}, \bibinfo{author}{EA, S.}, \bibinfo{author}{CP,
  L.}, \bibinfo{author}{Schuman, J.}, \bibinfo{author}{WG, S.},
  \bibinfo{author}{W, C.}, \bibinfo{author}{Hee, M.}, \bibinfo{author}{T, F.},
  \bibinfo{author}{K, G.}, \bibinfo{author}{CA, P.}, \bibinfo{author}{JG, F.},
  \bibinfo{year}{1991}.
\newblock \bibinfo{title}{Optical coherence tomography}.
\newblock \bibinfo{journal}{Science} \bibinfo{volume}{254},
  \bibinfo{pages}{1178}.
\bibitem[{Isola et~al.(2017)Isola, Zhu, Zhou and Efros}]{24}
\bibinfo{author}{Isola, P.}, \bibinfo{author}{Zhu, J.Y.},
  \bibinfo{author}{Zhou, T.}, \bibinfo{author}{Efros, A.},
  \bibinfo{year}{2017}.
\newblock \bibinfo{title}{Image-to-image translation with conditional
  adversarial networks}, pp. \bibinfo{pages}{5967--5976}.
\newblock \DOIprefix\doi{10.1109/CVPR.2017.632}.
\bibitem[{Jin et~al.(2019)Jin, Meng, Pham, Chen, Wei and Su}]{79}
\bibinfo{author}{Jin, Q.}, \bibinfo{author}{Meng, Z.}, \bibinfo{author}{Pham,
  T.D.}, \bibinfo{author}{Chen, Q.}, \bibinfo{author}{Wei, L.},
  \bibinfo{author}{Su, R.}, \bibinfo{year}{2019}.
\newblock \bibinfo{title}{Dunet: A deformable network for retinal vessel
  segmentation}.
\newblock \bibinfo{journal}{Knowledge-Based Systems} \bibinfo{volume}{178},
  \bibinfo{pages}{149–162}.
\newblock \URLprefix \url{http://dx.doi.org/10.1016/j.knosys.2019.04.025},
  \DOIprefix\doi{10.1016/j.knosys.2019.04.025}.
\bibitem[{Johnson et~al.(2016)Johnson, Alahi and Fei-Fei}]{39}
\bibinfo{author}{Johnson, J.}, \bibinfo{author}{Alahi, A.},
  \bibinfo{author}{Fei-Fei, L.}, \bibinfo{year}{2016}.
\newblock \bibinfo{title}{Perceptual losses for real-time style transfer and
  super-resolution}, pp. \bibinfo{pages}{694--711}.
\bibitem[{Kingma and Ba(2014)}]{73}
\bibinfo{author}{Kingma, D.}, \bibinfo{author}{Ba, J.}, \bibinfo{year}{2014}.
\newblock \bibinfo{title}{Adam: A method for stochastic optimization}.
\newblock \bibinfo{journal}{International Conference on Learning
  Representations} .
\bibitem[{Lee et~al.(2014)Lee, Xie, Gallagher, Zhang and Tu}]{69}
\bibinfo{author}{Lee, C.Y.}, \bibinfo{author}{Xie, S.},
  \bibinfo{author}{Gallagher, P.}, \bibinfo{author}{Zhang, Z.},
  \bibinfo{author}{Tu, Z.}, \bibinfo{year}{2014}.
\newblock \bibinfo{title}{Deeply-supervised nets}.
\newblock \href{http://arxiv.org/abs/1409.5185}{\tt arXiv:1409.5185}.
\bibitem[{{Ma} et~al.(2019){Ma}, {Wei}, {Ma}, {Shi} and {Zhu}}]{78}
\bibinfo{author}{{Ma}, J.}, \bibinfo{author}{{Wei}, M.}, \bibinfo{author}{{Ma},
  Z.}, \bibinfo{author}{{Shi}, L.}, \bibinfo{author}{{Zhu}, K.},
  \bibinfo{year}{2019}.
\newblock \bibinfo{title}{Retinal vessel segmentation based on generative
  adversarial network and dilated convolution}, in: \bibinfo{booktitle}{2019
  14th International Conference on Computer Science Education (ICCSE)}, pp.
  \bibinfo{pages}{282--287}.
\newblock \DOIprefix\doi{10.1109/ICCSE.2019.8845491}.
\bibitem[{Mao et~al.(2017)Mao, Li, Xie, Lau, Wang and Smolley}]{65}
\bibinfo{author}{Mao, X.}, \bibinfo{author}{Li, Q.}, \bibinfo{author}{Xie, H.},
  \bibinfo{author}{Lau, R.Y.K.}, \bibinfo{author}{Wang, Z.},
  \bibinfo{author}{Smolley, S.P.}, \bibinfo{year}{2017}.
\newblock \bibinfo{title}{Least squares generative adversarial networks}.
\newblock \href{http://arxiv.org/abs/1611.04076}{\tt arXiv:1611.04076}.
\bibitem[{Mirza and Osindero(2014)}]{23}
\bibinfo{author}{Mirza, M.}, \bibinfo{author}{Osindero, S.},
  \bibinfo{year}{2014}.
\newblock \bibinfo{title}{Conditional generative adversarial nets}.
\newblock \href{http://arxiv.org/abs/1411.1784}{\tt arXiv:1411.1784}.
\bibitem[{Owsley(2011)}]{4}
\bibinfo{author}{Owsley, C.}, \bibinfo{year}{2011}.
\newblock \bibinfo{title}{Aging and vision}.
\newblock \bibinfo{journal}{Vision Research} \bibinfo{volume}{51},
  \bibinfo{pages}{1610 -- 1622}.
\newblock \URLprefix
  \url{http://www.sciencedirect.com/science/article/pii/S0042698910005110},
  \DOIprefix\doi{https://doi.org/10.1016/j.visres.2010.10.020}.
  \bibinfo{note}{vision Research 50th Anniversary Issue: Part 2}.
\bibitem[{Pizer et~al.(1987)Pizer, Amburn, Austin, Cromartie, Geselowitz,
  Greer, {ter Haar Romeny}, Zimmerman and Zuiderveld}]{64}
\bibinfo{author}{Pizer, S.M.}, \bibinfo{author}{Amburn, E.P.},
  \bibinfo{author}{Austin, J.D.}, \bibinfo{author}{Cromartie, R.},
  \bibinfo{author}{Geselowitz, A.}, \bibinfo{author}{Greer, T.},
  \bibinfo{author}{{ter Haar Romeny}, B.}, \bibinfo{author}{Zimmerman, J.B.},
  \bibinfo{author}{Zuiderveld, K.}, \bibinfo{year}{1987}.
\newblock \bibinfo{title}{Adaptive histogram equalization and its variations}.
\newblock \bibinfo{journal}{Computer Vision, Graphics, and Image Processing}
  \bibinfo{volume}{39}, \bibinfo{pages}{355 -- 368}.
\newblock \URLprefix
  \url{http://www.sciencedirect.com/science/article/pii/S0734189X8780186X},
  \DOIprefix\doi{https://doi.org/10.1016/S0734-189X(87)80186-X}.
\bibitem[{Radford et~al.(2015)Radford, Metz and Chintala}]{27}
\bibinfo{author}{Radford, A.}, \bibinfo{author}{Metz, L.},
  \bibinfo{author}{Chintala, S.}, \bibinfo{year}{2015}.
\newblock \bibinfo{title}{Unsupervised representation learning with deep
  convolutional generative adversarial networks}.
\newblock \href{http://arxiv.org/abs/1511.06434}{\tt arXiv:1511.06434}.
\bibitem[{Ronneberger et~al.(2015)Ronneberger, Fischer and Brox}]{25}
\bibinfo{author}{Ronneberger, O.}, \bibinfo{author}{Fischer, P.},
  \bibinfo{author}{Brox, T.}, \bibinfo{year}{2015}.
\newblock \bibinfo{title}{U-net: Convolutional networks for biomedical image
  segmentation} .
\bibitem[{Son et~al.(2017)Son, Park and Jung}]{37}
\bibinfo{author}{Son, J.}, \bibinfo{author}{Park, S.J.}, \bibinfo{author}{Jung,
  K.H.}, \bibinfo{year}{2017}.
\newblock \bibinfo{title}{Retinal vessel segmentation in fundoscopic images
  with generative adversarial networks} .
\bibitem[{Tsai and Shah(1994)}]{70}
\bibinfo{author}{Tsai, P.S.}, \bibinfo{author}{Shah, M.}, \bibinfo{year}{1994}.
\newblock \bibinfo{title}{Shape from shading using linear approximation}.
\newblock \bibinfo{journal}{Image Vis. Comput.} \bibinfo{volume}{12},
  \bibinfo{pages}{487--498}.
\bibitem[{Vincent et~al.(2010)Vincent, Larochelle, Lajoie, Bengio and
  Manzagol}]{58}
\bibinfo{author}{Vincent, P.}, \bibinfo{author}{Larochelle, H.},
  \bibinfo{author}{Lajoie, I.}, \bibinfo{author}{Bengio, Y.},
  \bibinfo{author}{Manzagol, P.A.}, \bibinfo{year}{2010}.
\newblock \bibinfo{title}{Stacked denoising autoencoders: Learning useful
  representations in a deep network with a local denoising criterion}.
\newblock \bibinfo{journal}{J. Mach. Learn. Res.} \bibinfo{volume}{11},
  \bibinfo{pages}{3371–3408}.
\bibitem[{Voleti and Hubschman(2013)}]{3}
\bibinfo{author}{Voleti, V.B.}, \bibinfo{author}{Hubschman, J.P.},
  \bibinfo{year}{2013}.
\newblock \bibinfo{title}{Age-related eye disease}.
\newblock \bibinfo{journal}{Maturitas} \bibinfo{volume}{75}, \bibinfo{pages}{29
  -- 33}.
\newblock \URLprefix
  \url{http://www.sciencedirect.com/science/article/pii/S0378512213000273},
  \DOIprefix\doi{https://doi.org/10.1016/j.maturitas.2013.01.018}.
\bibitem[{Wang et~al.(2017)Wang, Xu, Wang and Tao}]{40}
\bibinfo{author}{Wang, C.}, \bibinfo{author}{Xu, C.}, \bibinfo{author}{Wang,
  C.}, \bibinfo{author}{Tao, D.}, \bibinfo{year}{2017}.
\newblock \bibinfo{title}{Perceptual adversarial networks for image-to-image
  transformation}.
\newblock \bibinfo{journal}{IEEE Transactions on Image Processing}
  \bibinfo{volume}{PP}.
\newblock \DOIprefix\doi{10.1109/TIP.2018.2836316}.
\bibitem[{Wang et~al.(2018)Wang, Dai, Zhao, Liu and Sun}]{59}
\bibinfo{author}{Wang, P.}, \bibinfo{author}{Dai, F.}, \bibinfo{author}{Zhao,
  M.}, \bibinfo{author}{Liu, X.}, \bibinfo{author}{Sun, J.},
  \bibinfo{year}{2018}.
\newblock \bibinfo{title}{Evaluation of retinal sfs reconstruction with oct
  images}, pp. \bibinfo{pages}{1--5}.
\newblock \DOIprefix\doi{10.1109/CISP-BMEI.2018.8633141}.
\bibitem[{Wang et~al.(2004)Wang, Bovik, Sheikh and Simoncelli}]{67}
\bibinfo{author}{Wang, Z.}, \bibinfo{author}{Bovik, A.},
  \bibinfo{author}{Sheikh, H.}, \bibinfo{author}{Simoncelli, E.},
  \bibinfo{year}{2004}.
\newblock \bibinfo{title}{Image quality assessment: From error visibility to
  structural similarity}.
\newblock \bibinfo{journal}{Image Processing, IEEE Transactions on}
  \bibinfo{volume}{13}, \bibinfo{pages}{600 -- 612}.
\newblock \DOIprefix\doi{10.1109/TIP.2003.819861}.
\bibitem[{Yang et~al.(2017)Yang, Yan, Zhang, Yu, Shi, Mou, Kalra, Zhang, Sun
  and Wang}]{35}
\bibinfo{author}{Yang, Q.}, \bibinfo{author}{Yan, P.}, \bibinfo{author}{Zhang,
  Y.}, \bibinfo{author}{Yu, H.}, \bibinfo{author}{Shi, Y.},
  \bibinfo{author}{Mou, X.}, \bibinfo{author}{Kalra, M.},
  \bibinfo{author}{Zhang, Y.}, \bibinfo{author}{Sun, L.},
  \bibinfo{author}{Wang, G.}, \bibinfo{year}{2017}.
\newblock \bibinfo{title}{Low-dose ct image denoising using a generative
  adversarial network with wasserstein distance and perceptual loss}.
\newblock \bibinfo{journal}{IEEE Transactions on Medical Imaging}
  \bibinfo{volume}{PP}.
\newblock \DOIprefix\doi{10.1109/TMI.2018.2827462}.
\bibitem[{Zhang et~al.(2020)Zhang, Pan, Zhang and Tiong}]{80}
\bibinfo{author}{Zhang, G.}, \bibinfo{author}{Pan, Y.}, \bibinfo{author}{Zhang,
  L.}, \bibinfo{author}{Tiong, R.L.K.}, \bibinfo{year}{2020}.
\newblock \bibinfo{title}{Cross-scale generative adversarial network for crowd
  density estimation from images}.
\newblock \bibinfo{journal}{Engineering Applications of Artificial
  Intelligence} \bibinfo{volume}{94}, \bibinfo{pages}{103777}.
\newblock \URLprefix
  \url{http://www.sciencedirect.com/science/article/pii/S0952197620301743},
  \DOIprefix\doi{https://doi.org/10.1016/j.engappai.2020.103777}.
\bibitem[{Zhang et~al.(2017)Zhang, Sindagi and Patel}]{75}
\bibinfo{author}{Zhang, H.}, \bibinfo{author}{Sindagi, V.},
  \bibinfo{author}{Patel, V.}, \bibinfo{year}{2017}.
\newblock \bibinfo{title}{Image de-raining using a conditional generative
  adversarial network}.
\newblock \bibinfo{journal}{IEEE Transactions on Circuits and Systems for Video
  Technology} \bibinfo{volume}{PP}.
\newblock \DOIprefix\doi{10.1109/TCSVT.2019.2920407}.
\bibitem[{Zhang et~al.(2018)Zhang, Isola, Efros, Shechtman and Wang}]{68}
\bibinfo{author}{Zhang, R.}, \bibinfo{author}{Isola, P.},
  \bibinfo{author}{Efros, A.}, \bibinfo{author}{Shechtman, E.},
  \bibinfo{author}{Wang, O.}, \bibinfo{year}{2018}.
\newblock \bibinfo{title}{The unreasonable effectiveness of deep features as a
  perceptual metric} .
\bibitem[{Zhao et~al.(2019)Zhao, Zhang, Li, Hwang, Gao, Fang, Jiang and
  Huang}]{76}
\bibinfo{author}{Zhao, J.}, \bibinfo{author}{Zhang, J.}, \bibinfo{author}{Li,
  Z.}, \bibinfo{author}{Hwang, J.N.}, \bibinfo{author}{Gao, Y.},
  \bibinfo{author}{Fang, Z.}, \bibinfo{author}{Jiang, X.},
  \bibinfo{author}{Huang, B.}, \bibinfo{year}{2019}.
\newblock \bibinfo{title}{Dd-cyclegan: Unpaired image dehazing via
  double-discriminator cycle-consistent generative adversarial network}.
\newblock \bibinfo{journal}{Engineering Applications of Artificial
  Intelligence} \bibinfo{volume}{82}, \bibinfo{pages}{263 -- 271}.
\newblock \URLprefix
  \url{http://www.sciencedirect.com/science/article/pii/S0952197619300806},
  \DOIprefix\doi{https://doi.org/10.1016/j.engappai.2019.04.003}.
\bibitem[{Zhou et~al.(2018)Zhou, Siddiquee, Tajbakhsh and Liang}]{21}
\bibinfo{author}{Zhou, Z.}, \bibinfo{author}{Siddiquee, M.M.R.},
  \bibinfo{author}{Tajbakhsh, N.}, \bibinfo{author}{Liang, J.},
  \bibinfo{year}{2018}.
\newblock \bibinfo{title}{Unet++: A nested u-net architecture for medical image
  segmentation}.
\newblock \href{http://arxiv.org/abs/1807.10165}{\tt arXiv:1807.10165}.

\end{thebibliography}

\end{document}